# Dual Deep Learning Approach for Non-invasive Renal Tumour Subtyping with VERDICT-MRI

*Snigdha Sen[1], Lorna Smith[2], Lucy Caselton[3], Joey Clemente[2], Maxine Tran[4], Shonit Punwani[2], David Atkinson[2], Richard L Hesketh[3]\* and Eleftheria Panagiotaki[1]\**

1. UCL Hawkes Institute, 90 High Holborn, London, UK
2. UCL Centre for Medical Imaging, Charles Bell House, London, UK
3. UCL Department of Imaging, London, UK
4. UCL Surgery and Interventional Science, London, UK

**Corresponding author:** Snigdha Sen; snigdha.sen.20@ucl.ac.uk

## Abstract

**Purpose**

To characterise renal tumour microstructure using diffusion MRI (dMRI); via the Vascular, Extracellular and Restricted Diffusion for Cytometry in Tumours (VERDICT)-MRI framework with self-supervised learning. Additionally, to identify a clinically feasible acquisition protocol for further studies.

**Materials and Methods**

Comprehensive dMRI datasets were acquired from 14 patients with 17 renal tumours (15 biopsy-confirmed), with nine b-values in the range $b=[0,2500]s/mm^2$. A three-compartment VERDICT model for renal tumours was fitted to the dMRI data using a self-supervised deep neural network and analyse ROIs drawn by an experienced uroradiologist. Statistical differences between groups were computed using the Wilcoxon's signed-rank test. Lastly, an economical acquisition protocol for future studies with larger patient cohorts was optimised using a recursive feature selection approach.

**Results**

The VERDICT model described the diffusion data in renal tumours more accurately than IVIM or ADC. Combined with self-supervised deep learning, VERDICT identified significant differences in the intracellular volume fraction between cancerous and normal tissue, and in the vascular volume fraction between vascular and non-vascular. The feature selector yields a 4 b-value acquisition b = [70, 150, 1000, 2000], with a duration of 14 minutes.

**Conclusion**

VERDICT-MRI combined with self-supervised learning characterises RCC subtypes, revealing microstructure traits that are not accessible via ADC. An optimal protocol for renal tumours which has a clinically feasible acquisition duration has been identified using deep-learning-based feature selection.

**Summary**

A VERDICT-MRI diffusion MRI model has been developed to characterise renal tumours with deep learning, demonstrating improved description of the underlying microstructure than ADC and IVIM models, with a clinically viable acquisition.

**Key Points**

- The VERDICT intracellular volume fraction is significantly higher in cancerous than normal kidney tissue.
- The VERDICT vascular volume fraction is statistically higher in vascular renal tumour subtypes than non-vascular with the potential to eliminate the need for contrast-enhanced renal imaging.
- A dual-network approach identifies an optimal protocol for renal microstructure characterisation, with an acquisition time reduction of over 30 minutes.

**Introduction**

Renal cancer accounts for roughly 2% of all cancer diagnoses and deaths worldwide [1]. Over 90% of primary renal cancers are renal cell carcinoma (RCC), arising from the renal tubule epithelium [2]. The 2022 WHO classification subdivides RCC into 21 different histotypes of which the commonest are clear-cell RCC (ccRCC) (~70%), papillary RCC (pRCC) (~10%), chromophobe (chRCC) (~5%) [3]. Benign tumours of the kidney can also occur, notably oncocytoma and angiomyolipoma [4]. The main diagnostic imaging for renal tumours is computed tomography (CT), which can diagnose RCC with a sensitivity of over 95% and a specificity of around 90% [5]. While different tumour subtypes exhibit typical imaging features on both CT and MRI, there is considerable overlap between certain types e.g., oncocytoma, clear cell and chRCC [6]. With the exception of the identification of "macroscopic fat" in lipid-rich angiomyolipomas, there are currently no imaging biomarkers with sufficient accuracy to differentiate benign and malignant tumours in clinical practice [7]. A historical reluctance to use biopsy due to the risks, sampling error and a potentially low negative predictive value have meant that rates of benign histology following nephrectomy remain as high as 31% [8].

Therefore, new imaging biomarkers are required that are capable of accurately defining both lesion subtype and metastatic risk. MRI techniques offer valuable insights into tumour biology; particularly diffusion-weighted (DW) MRI, which can access restricted diffusion via the apparent diffusion coefficient (ADC). DW-MRI has been used to differentiate RCC subgroups [9-11] and has proven useful in characterising indeterminate small renal lesions, whether inflammatory or malignant, as both may exhibit restricted diffusion [12]. However, whilst most clinical DW-MRI studies focus on the ADC, it conflates multiple physiological parameters, reducing its specificity for identifying disease mechanisms, with studies showing varying performance in RCC subtype characterisation [13-15]. The intravoxel incoherent motion (IVIM) model, which separates diffusion in tissue from pseudo-diffusion in blood vessels, has demonstrated enhanced performance by capturing tumour vascularity [16-18]. While IVIM improves upon the uni-compartmental ADC, it still oversimplifies the complex underlying tissue microstructure.

The Vascular, Extracellular and Restricted Diffusion for Cytometry in Tumours (VERDICT)-MRI framework combines a specific diffusion acquisition protocol with a three-compartment biophysical model of the DW-MRI signal in tumour tissue [19]. Previously, VERDICT has shown effective tumour characterisation in colorectal [20], prostate [21,22] and brain tumours [23]. Specifically, in prostate cancer where the problem of lesion over-treatment is similar to that in renal tumours, VERDICT improves differentiation between benign and low-grade malignant lesions compared to the ADC [24,25]. In this work, we trial VERDICT-MRI for renal tumours for the first time, in a cohort of 14 patients with 17 renal masses. The patients participated in a specific DW acquisition to obtain rich datasets for modelling analysis, and a kidney-specific VERDICT model was identified and fitted to this data using a self-supervised deep learning method [26]. The acquisition of comprehensive datasets took 40 min; a clinically impractical acquisition time. Traditionally, optimised protocols are identified via the Fisher information, a computationally expensive and model-restricted approach [27]. We investigated recursive feature elimination methods that have been used previously to identify optimal VERDICT imaging protocols [28,29], with a model-free approach adaptable to different tasks.

We found that our VERDICT model for renal tissue outperformed ADC and IVIM at describing the diffusion-weighted data and underlying microstructure. The VERDICT intracellular volume fraction was significantly higher in cancer than normal kidney tissue and revealed similar trends to those predicted by histology, whilst the vascular volume fraction was statistically higher in vascular tumours than in non-vascular. Finally, our dual-network feature selection method identified an economical acquisition protocol for clinical use, reducing the scan time by 36 minutes and maintaining diagnostic power comparable to the full protocol.

## Materials and Methods

### Patient Cohort

The study was approved by the UCL Research Ethics Committee (07/Q0502/15) and all patients provided informed written consent. Patients were eligible to participate if they had a diagnosis of a T1 solid renal tumour made on standard of care imaging, were able to have an MRI and had not had a recent biopsy within the last three months. Patients with renal tumours were identified at the Royal Free Hospital Renal Cancer Supra-Network Multi-Disciplinary Team Meeting (SMDT).

Fourteen patients with 17 tumours (15 of which had histological confirmation and were included in the analysis) underwent MRI examination. The patients had a mean age of 66 ± 11 years, and a 1:3.6 female:male ratio (Table 1).

| Tumour type | Total | WHO/ISUP grades |
|---|---|---|
| ccRCC | 5 | 1 (n = 3), 2 (n =2) |
| pRCC | 2 | 1 (n = 1), 3 (n = 1) |
| Unclassified RCC (uRCC) | 1 | 3 (n = 1) |
| chRCC* | 4 | n/a |
| Classical oncocytoma | 3 | n/a |

**Table 1.** Tumour types included in the study. *Includes low grade oncocytic neoplasm.

### VERDICT-MRI Acquisition

VERDICT-MRI was performed on a 3T MRI system (Ingenia; Philips, Best, Netherlands), using a 16-channel body coil (SENSE XL Torso, Philips, Best, Netherlands), using a pulsed gradient spin-echo (PGSE) sequence with echo-planar imaging (EPI) readout in the coronal imaging plane. The acquisition was informed by the optimised protocol for prostate [30], with additional measurements to obtain rich datasets for analysis and emphasis placed on lower b-values to reflect the vascularity of renal microstructure. The imaging parameters were as follows: repetition time (TR), 2000–3349ms; field of view, 220 x 200 mm; voxel size, 1.25 x 1.25 x 5 mm; no interslice gap; acquisition matrix, 176 x 176. The VERDICT acquisition protocol for kidney is: $b$ = [70, 90, 150, 500, 1000, 1500, 2000, 2200, 2500] s/mm$^2$; $\delta$ = [4.8, 4.8, 4.8, 12.0, 12.0, 26.3, 16.8, 16.8, 21.4] ms; $\Delta$ = [27.0, 27.0, 27.0, 34.0, 34.0, 47.0, 37.5, 37.5, 43.5] ms. For each of the nine combinations of $b/\delta/\Delta$ we used the minimum possible echo time (TE), giving TEs of 54–87ms, and we acquired a separate $b = 0$ image for each TE, resulting in 18 image volumes, each with 14 slices. The acquisition time was approximately 40 min.

**Image Analysis**

The preprocessing pipeline included denoising using MP-PCA [31] (MrTrix3 'dwidenoise' [32]), and correction for Gibbs ringing [33]. We applied mutual-information rigid and affine registration [34] and divided the DW-MRI volumes by their matched b=0 for normalisation. Regions of interest (ROIs) for tumour, simple cysts and normal renal parenchyma were drawn manually in ITKSnap v4.0 by a board-certified study radiologist using co-registered T2-weighted images [35]. Histological subtype and tumour WHO / ISUP (International Society of Urological Pathology) grade was determined from either lesion biopsy or surgery and compared to imaging tumour analysis. 10 unpaired surgical histological samples were of the main tumour types (ccRCC, pRCC, chRCC and oncocytoma) were stained with haematoxylin-and-eosin (H&E) and CD34 (a vascular endothelial marker) immunohistochemistry. Slides were scanned and the digital images analysed using the software QuPath [36]. Cell density and vascularity were used as histological validation of the VERDICT model.

**VERDICT Model**

The VERDICT model for kidney has three compartments that characterise the diffusion of water molecules in the intracellular (IC), vascular (VASC) and extracellular-extravascular space (EES) in tumours, with no exchange between the components [19,20].
The total normalised signal is:
$$\frac{S}{S_0} = f_{VASC} S_{VASC}(d_{VASC}, b) + f_{IC} S_{IC}(d_{IC}, R, b, \Delta, \delta) + f_{EES} S_{EES}(d_{EES}, b) \qquad (1)$$
where $f_i$ is the volume fraction and $S_i$ is the normalised signal from water molecules in population $i$, where $i$ = IC, VASC or EES. The vascular signal fraction, $f_{VASC}$, is computed as $1 - f_{IC} - f_{EES}$, since $\sum_{i=1}^{3} f_i = 1$ and $0 \leq f_i \leq 1$, and $S_0$ is the signal with no diffusion weighting. Here $b$ is the b-value, $\Delta$ is the gradient pulse separation and $\delta$ is the gradient pulse duration.

To represent renal tissue, the IC compartment was modelled as an impermeable sphere of radius $R$ with IC diffusivity fixed at $d_{IC} = 2\ \mu m^2/ms$ and the EES compartment as Gaussian free diffusion with diffusivity $d_{EES} = 2\ \mu m^2/ms$. The vascular compartment was randomly-oriented sticks with pseudo-diffusivity $d_{VASC} = 50\ \mu m^2/ms$ to reflect the high vascularity of kidney tissue (see Supplementary Fig. 1). Three free parameters were estimated: $f_{IC}$ (IC volume fraction), $f_{EES}$ (EES volume fraction) and cell radius R within biophysically-realistic parameter ranges:

**Model Fitting**

The renal VERDICT model is fit to the data using self-supervised deep learning, previously used in [26] with the prostate model. These techniques remove the need for training data, instead extracting labels from the input data itself. They work to estimate the model parameters and reconstruct the signal, then calculate the training loss as the mean-squared error between the predicted signal, S^, and the input signal S [37] to find the optimal estimates. A fully connected neural network with three hidden layers, each with 18 neurons (i.e. the number of image volumes, a spherically-averaged b-value image with a corresponding b0), was implemented using *PyTorch* 1.12.1. The output layer is fed into the VERDICT model equation to generate the predicted signal S^. We used the ADAM optimiser, early stopping to mitigate overfitting, and a learning rate of 0.0001. We used dropout of p=0.5 and constrained the parameter values to the ranges above using the *PyTorch* clamp function. Training and prediction for each masked dMRI data set took about 50 s per subject.

**Protocol Optimisation**

The protocol mentioned above provides rich DW datasets useful for model development and analysis, however the acquisition time of ~40 minutes is not clinically feasible. We develop a deep-learning approach based on similar work in literature [28,29,39] to select the most informative subset from the full protocol of nine b-values.

Features were selected from 27 measurements from the DW images acquired along the three orthogonal directions per b-value. The non-DW measurements were excluded. Each DW dataset was normalised and fed into the feature selection feed-forward neural network. This estimated importance scores for the 27 measurements – it comprised an input layer of 27 nodes, a hidden layer with 64 nodes (followed by ReLU and batch normalisation) and an output layer with 12 nodes and a sigmoid activation function to ensure feature scores in [0,1]. Using the subset of the measurements with the highest scores, the predictive network was trained to approximate the targets, consisting of an input layer of 12 nodes, a hidden layer of 64 nodes and ReLU activation and an output layer of size 27, matching the number of target measurements.

The two networks were trained in tandem for 100 epochs and the loss was computed as the MSE between the predicted and ground truth target values, which were the input 27 measurements. A learning rate of $1e^{-5}$ was used, with the ADAM optimiser and dropout to prevent overfitting. Upon training completion, we extract the final optimal feature set by computing the average importance scores across all subjects and extracting a final protocol of 4 b-values. The method is outlined in Fig. 1.

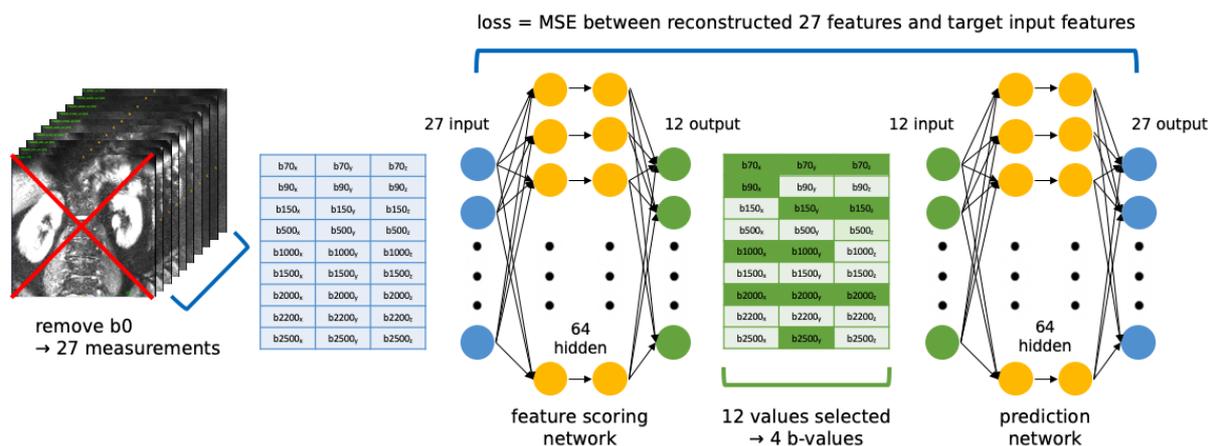

**Figure 1:** Schematic outlining the protocol optimisation dual-network strategy. We first remove the non-DW images from the image volumes, resulting in 27 measurements per patient. Our feature scoring network takes in the 27 measurements (27 input nodes), has hidden layers with 64 nodes and outputs 12 measurements. These are then used to reconstruct the target initial 27 measurements in the predictor network, and the loss is computed as the MSE between the output and the target. We train on 12 of the patients for 100 epochs, and test on the remaining three, choosing b = 70, 150, 1000, 2000 as our final protocol. This gives a reduction in acquisition time of 34 minutes.

**Statistical Analysis**

Data were analysed using Python 3.11 (SciPy 1.15.0 [38]) and statistical tests performed were Wilcoxon's signed-rank tests with errors reported as standard deviation, unless stated otherwise. *p*-values are summarised in figures as: <0.0001, ****; 0.0001 – 0.001, ***; 0.001 – 0.01, **; 0.01 to 0.05, *.

## Results

### Histological Validation of the VERDICT Model

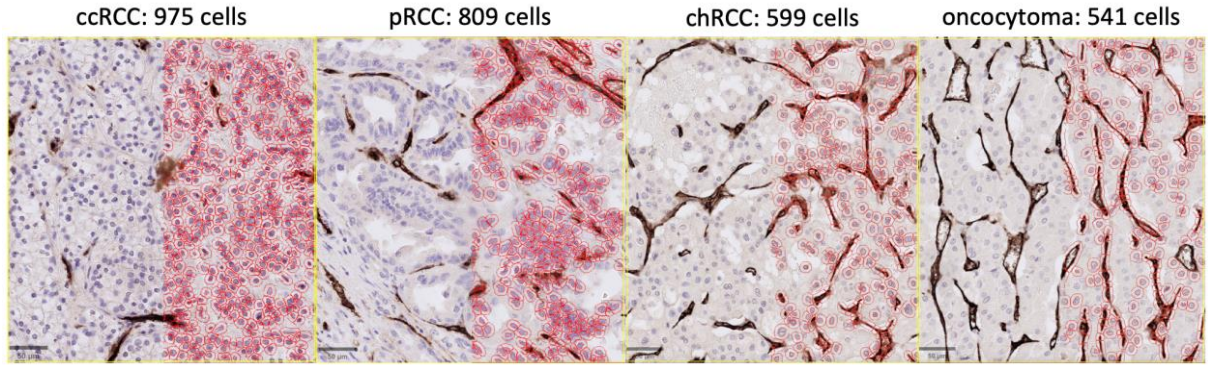

**Figure 2:** Histological analysis of cell size using H&E-staining. We observe that ccRCC and pRCC have a higher number of cells in the same area (180,000$\mu m^2$) of tumour in comparison to chRCC and oncocytoma, which agrees with literatures that they are more cellular tumour types.

ccRCC and pRCC demonstrated higher cell density in comparison to chRCC and oncocytoma (Fig. 2). Vascularity, measured by CD34 staining, grouped tumours into a vascular group of tumours (ccRCC, 12.6% positivity; oncocytoma, 10.4%; and chRCC, 7.1%) and non-vascular tumours (pRCC, 2.8%).

### VERDICT, IVIM and ADC Model Fitting of Normal and Cancerous Tissue

The VERDICT, IVIM and ADC models were fitted to the data from ROIs in tumour and normal kidney in Fig. 3. In both low and high-grade renal tumours, the MSE of the VERDICT fit is lower than ADC and IVIM in both normal and cancerous tissue, suggesting that it describes the data most accurately.

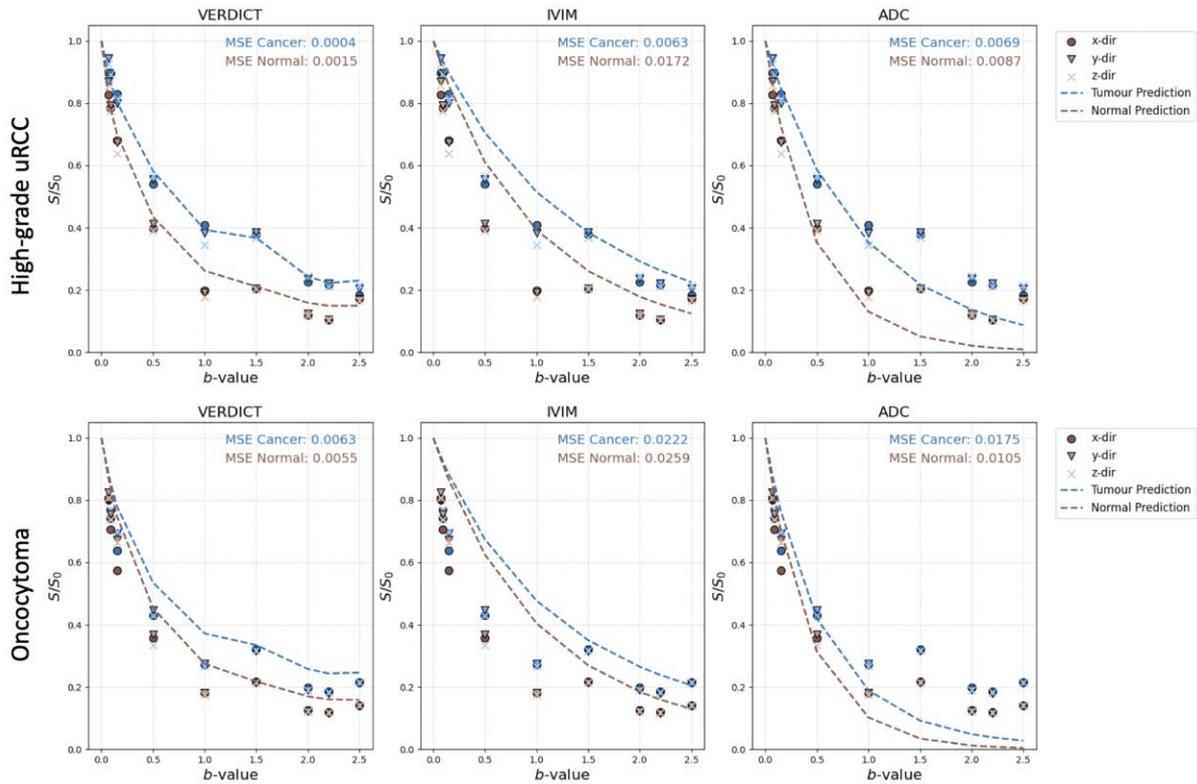

**Figure 3:** Signal prediction vs DWI data for tumour and normal renal tissue via VERDICT vs. IVIM vs. ADC for a high-grade tumour and benign oncocytoma. The DW signal data are plotted as the points, the curve represents the model's signal prediction and the MSE is calculated as the difference between the true and predicted value.

**VERDICT and ADC Parameter Estimates**

Parametric maps were generated for the VERDICT parameters ($f_{IC}$, $f_{EES}$, $f_{VASC}$ and R) and ADC, for each of the patient groupings (Fig. 4). $f_{IC}$ and ADC ranged between 0.06 and 0.21 (SD: 0.03) and 2.05 and 2.76 (SD: 0.2), respectively, for benign renal parenchyma across all patients. $f_{IC}$ discriminated between cancerous from normal tissue (0.32 vs. 0.12; $p < 0.01$), as did ADC (1.6 vs. 2.3; $p < 0.05$) (Figure 5a and b). The $f_{IC}$ for oncocytomas (benign) was not significantly different to normal tissue (0.17 vs. 0.12; p = n.s.). There were trends towards higher $f_{IC}$ and R in the high-grade uRCC ($f_{IC}$ = 0.59) and pRCC ($f_{IC}$ = 0.55) compared to low grade tumours ($f_{IC}$ = 0.21 ± 0.07) (Figs. 4 and 5). A similar trend was seen with ADC.

The tumours were grouped into vascular (ccRCC, oncocytoma and chromophobe tumours), and non-vascular (uRCC, pRCC) based on the unpaired histological analysis and degree of enhancement demonstrated on nephrographic-phase CT for each patient.

The $f_{VASC}$ was significantly higher in the vascular tumours than in non-vascular (mean ± SD: 0.29 ± 0.05 vs. 0.13 ± 0.02, p < 0.05; Fig. 5c).

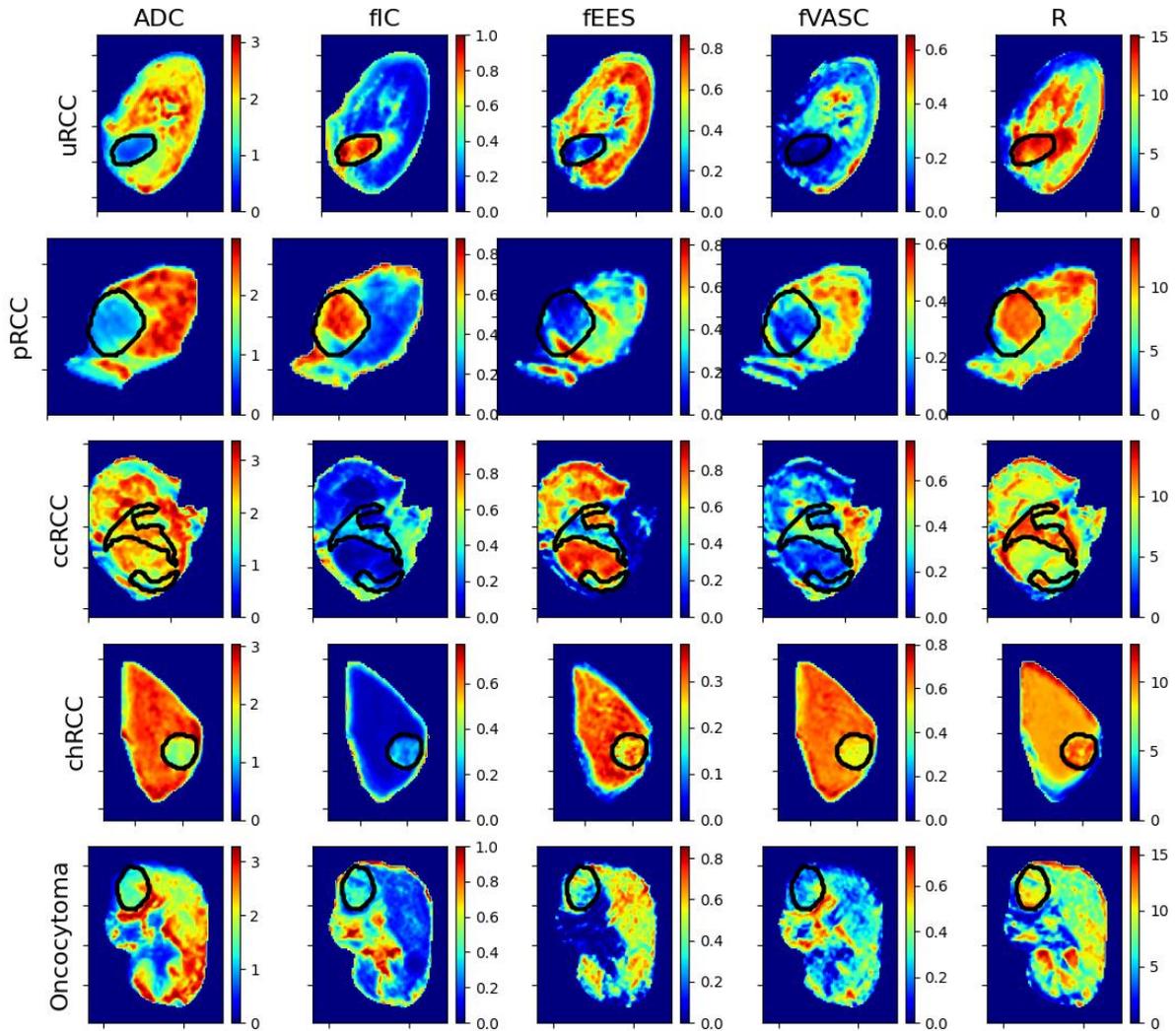

**Figure 4:** Parameter maps for ADC and VERDICT parameters in all patient groups. We observe higher $f_{IC}$ and R for more cellular tumours, (uRCC, pRCC and ccRCC), and elevated $f_{VASC}$ in more vascular masses (ccRCC, oncocytoma). ADC is lower for uRCC, pRCC and chRCC, but not for the other groups.

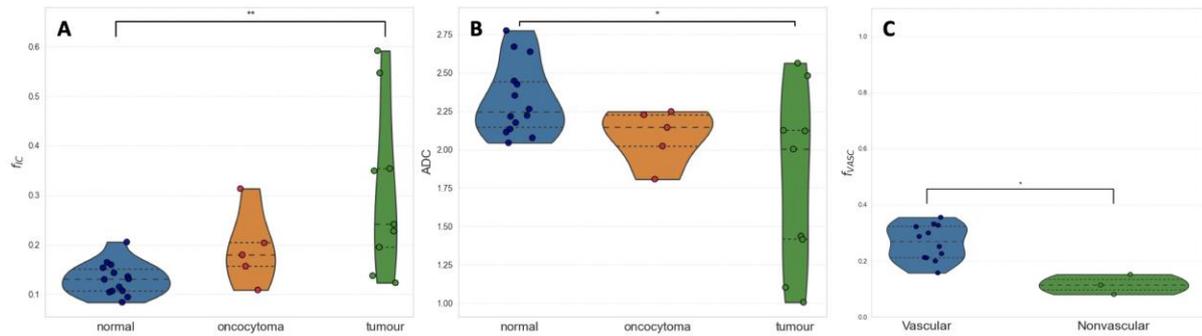

**Figure 5**: (A) VERDICT $f_{IC}$ and (B) ADC in normal kidney tissue, benign oncocytomas and cancerous tumours. The $f_{IC}$ can discriminate cancer from normal tissue ($p < 0.01$), as can ADC with $p < 0.05$. (C) VERDICT $f_{VASC}$ in vascular vs. non-vascular tumours – the $f_{VASC}$ can discriminate these ($p < 0.05$).

**Deep Learning Protocol Optimisation**

The deep learning feature selection method produced a protocol of four optimally-informative b-values: 70, 150, 1000, and 2000s/mm$^2$. In Fig. 6, we reproduce our results with this protocol, demonstrating the quality of the parameter maps is largely preserved, with some smoothing. We plot the differences between these maps and the original and find that they are generally small with more variation outside of the tumour. We also find the MSE between the signal predictions and DW data remains similar to Fig. 3 in a benign and cancerous tumour. In Supplementary Fig. 3, we recreate Fig. 5 with the optimised protocol, finding similar statistical significance to the full protocol. The optimised protocol has an acquisition time of ~14 minutes, a reduction of over 30 minutes from the full protocol.

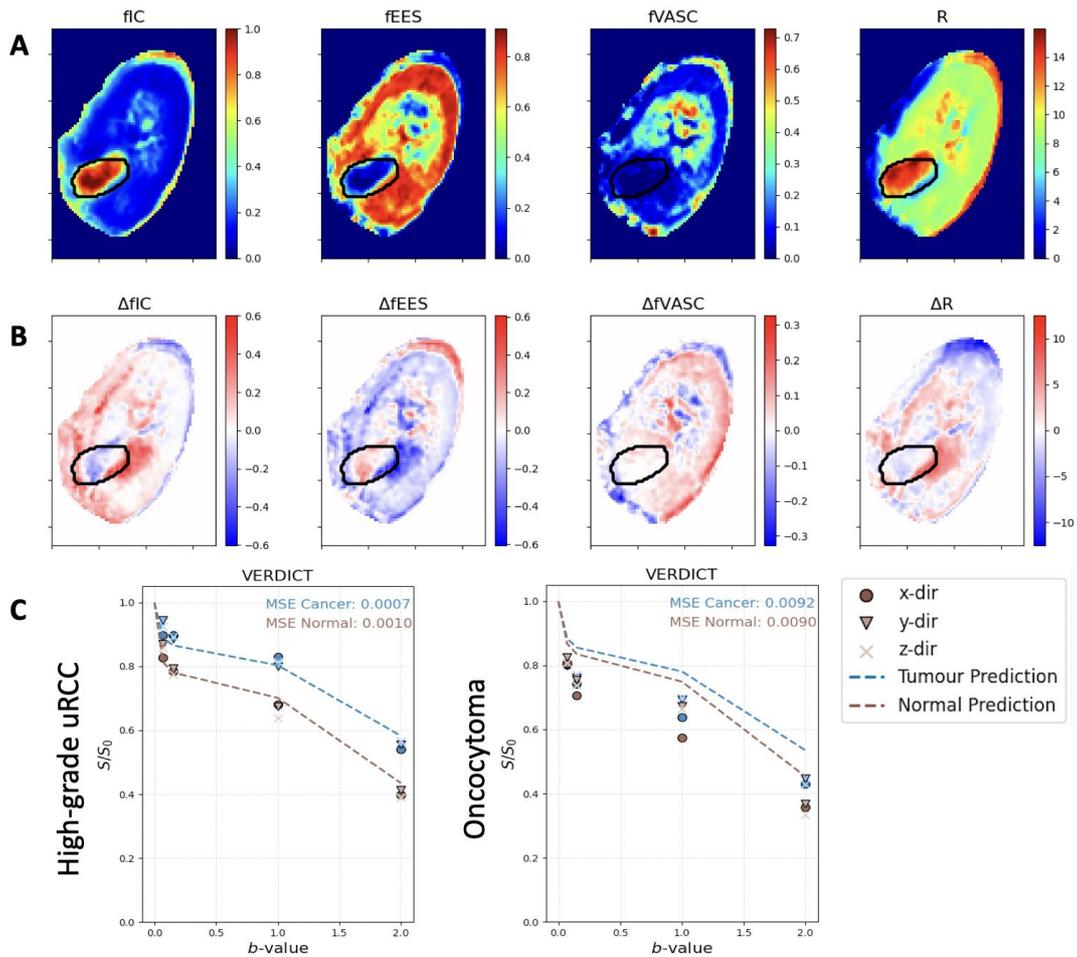

**Figure 6.** In (A) we show the maps produced using the identified shorter protocol, and in (B) we plot the difference between these maps and the original. We observe that they maintain the map features with some smoothing, and more variation is seen outside the tumour region. In (B) we plot the signal prediction vs. DWI data for a high-grade tumour and benign oncocytoma, observing that the MSE between the data and predictions remains similar to that seen with the full protocol.

## Discussion

In this work, we create a model for renal tumours within the VERDICT-MRI framework for the first time, in a cohort of 14 patients. We acquire rich datasets of nine b-values to develop and fit a three-compartment VERDICT model for the kidney using self-supervised deep learning. We show that VERDICT provides a better description of the data than IVIM and ADC and obtain statistically significant differences between tissue types in the intracellular and vascular volume fractions. Finally, we find that our results hold when using a significantly shortened protocol of four b-values, demonstrating that this technique can be translated to the clinic with ease.

In Fig. 3, we show that the predicted signal from VERDICT matches the DW data more closely in terms of MSE than IVIM or ADC in both tumour and benign tissue. This suggests VERDICT provides a better description of the underlying microstructure, agreeing with previous studies where VERDICT outperforms both models in the prostate [19,22]. We note the MSE is lower in tumour regions than benign; we hypothesise that this is because the model is developed to describe the DW signal in tumour tissue specifically, therefore describes those features more accurately.

Histologically, ccRCC and pRCC demonstrated higher cellularity than chRCC and benign oncocytoma (Fig. 4), while chRCC and oncocytoma appear similar, explaining the challenge of differentiating these tumours on imaging. This also agrees with our results in Supplementary Fig. 2, where we note that for both $f_{IC}$ and ADC, our results overlap considerably for chRCC and oncocytoma. The higher $f_{IC}$ estimates in cancerous tumours, particularly the high-grade tumours, compared to the normal renal parenchyma agrees with prior knowledge that cancerous tumours are typically hypercellular [40]. Benign oncocytomas could not be discriminated from either normal tissue or cancer at a statistically significant level with either $f_{IC}$ or ADC however, highlighting their confounding nature.

The unique microstructural information available via VERDICT compared to ADC permitted comparison of tumour vascularity. Informed by our prior knowledge from CT imaging we grouped tumours into a vascular and non-vascular group based on their CT-enhancement characteristics and showed the $f_{VASC}$ to be significantly higher in the vascular tumours group. Inferring enhancement from DWI and VERDICT could potentially avoid contrast administration clinically.

Lastly, in Fig. 5 we recreate our results with the optimised protocol identified via deep-learning-based feature selection and show consistency with those achieved with the full protocol. Along with Supplementary Fig. 3, these results demonstrate that similar diagnostic performance can be achieved whilst reducing the acquisition time by over 30 minutes, enabling the clinical translation of this work. This result also motivates the development of VERDICT models for other organs, as we demonstrate how the use of a few comprehensive datasets for model development can be used to identify clinical acquisitions.

This work is limited primarily by the small patient cohort, resulting in few patients per tumour subtype, which limits the statistical significance of these results. This is in part due to the long acquisition protocol; with the identified economical protocol, we hope to soon be able to trial VERDICT on a much larger number of kidney patients with a variety of subtypes. Additionally, whilst some histological analysis has been conducted, these are not paired samples with the imaging data; future studies will include paired histology and DWI for more in-depth analysis and quantitative results.

In conclusion, this work develops a VERDICT-MRI model for renal tumours, and analyses results in a cohort of 14 patients. We show that VERDICT provides better data description and enhanced microstructural information over ADC, such as differences between patient groups in intracellular and vascular volume fractions. We exploit novel deep learning advances for both the model fitting approach to ensure robustness, and to identify a significantly shorter acquisition protocol. We find that we can extract equivalent diagnostic performance with our abbreviated protocol as with the original, allowing for smooth clinical translation of the technique.

**Supplementary Materials**

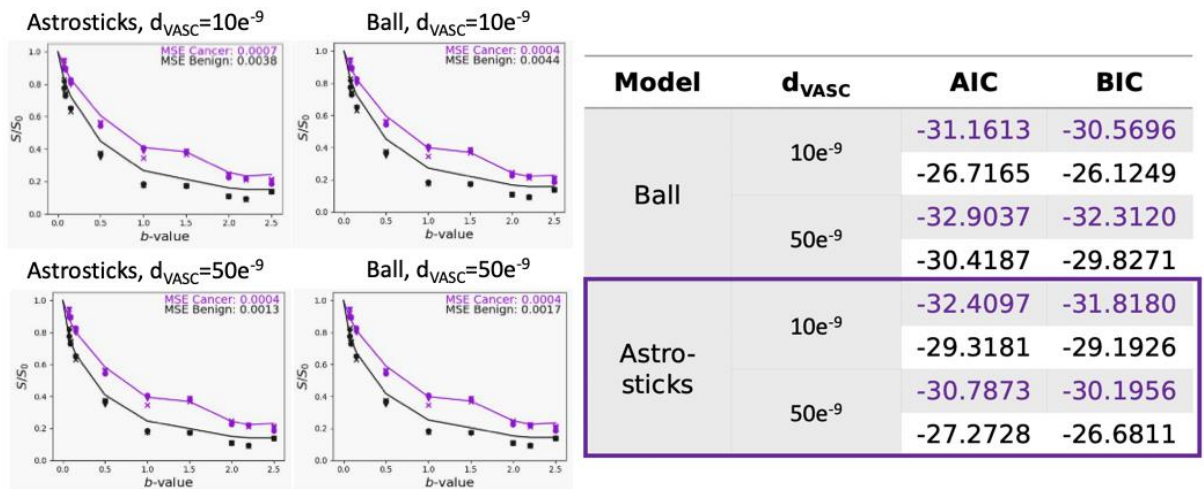

**Supplementary Figure 1:** Modelling analysis to develop a VERDICT model for renal tissue. We experimented with the vascular compartment, trialling both a 'ball' and 'astrosticks' geometry with $d_{VASC} = 10\ \mu m^2/ms$ and $d_{VASC} = 50\ \mu m^2/ms$. We found that the 'astrosticks' compartment with $d_{VASC} = 50\ \mu m^2/ms$ was the best fit to the DW data and also had the lowest Akaike's Information Criterion and Bayesian Information Criterion.

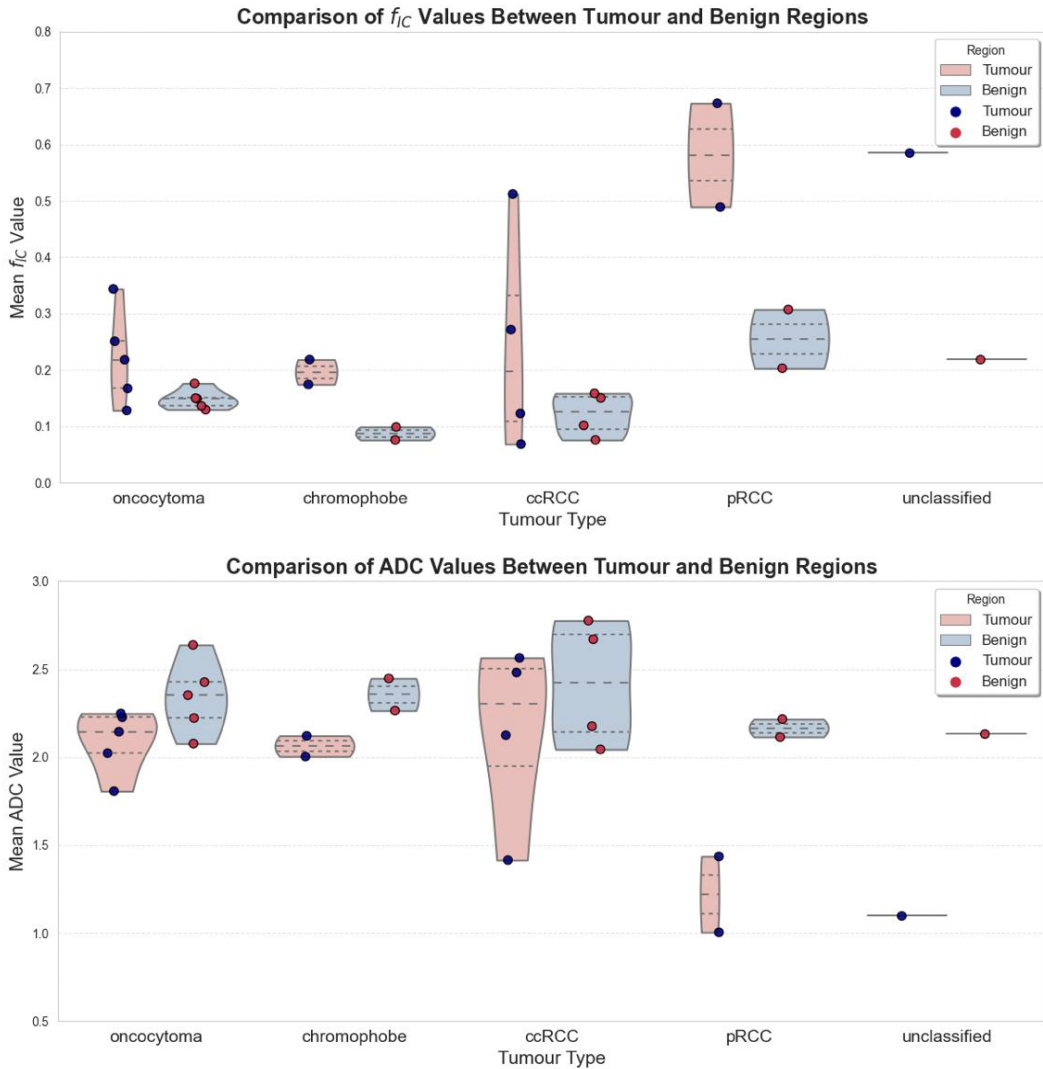

**Supplementary Figure 2:** Violin plots showing (A) $f_{IC}$ and (B) ADC estimates in patient groups. We observe good separation between tumour and healthy tissue for chRCC, pRCC and uRCC with both parameters, but not for the other groups. Particularly, for both groups, there is a large spread in tumour estimates for ccRCC, possibly due to the range of tumour grades in our cohort. The plots suggest that oncocytoma and chRCC have similar traits, which is supported by the histology in Fig. 2, whilst pRCC and uRCC have higher $f_{IC}$/lower ADC. The parameter estimates in tumours and benign tissue for each group are more consistent for $f_{IC}$ than for ADC (i.e. less spread).

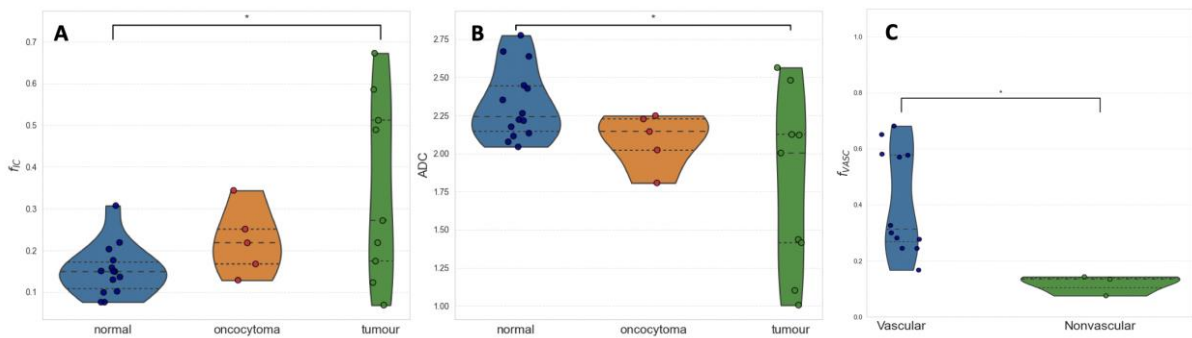

**Supplementary Figure 3:** Recreations of the violin plots in Fig. 5, with the reduced protocol identified via the feature selection network. We observe that we achieve statistical significance with $f_{IC}$ at $p < 0.05$ and the same statistical significance with the vascular volume fraction as we obtained with the full protocol. This demonstrates that we do not lose significant microstructural information with our significantly shorter acquisition.

## References

1. Padala SA, Barsouk A, Thandra KC, Saginala K, Mohammed A, Vakiti A, et al. Epidemiology of Renal Cell Carcinoma. *World J Oncol*. 2020 Jun;11(3):79-87. doi: 10.14740/wjon1279.
2. Hsieh JJ, Purdue MP, Signoretti S, Swanton C, Albiges L, Schmidinger M, et al. Renal cell carcinoma. *Nat Rev Dis Primers*. 2017 Mar 9;3:17009. doi: 10.1038/nrdp.2017.9.
3. Lobo J, Ohashi R, Amin MB, Berney DM, Compérat EM, Cree IA, et al. (2022), WHO 2022 landscape of papillary and chromophobe renal cell carcinoma. *Histopathology*, 81: 426-438. doi: 10.1111/his.14700
4. Tamboli P, Ro JY, Amin MB, Ligato S, Ayala AG. Benign tumors and tumor-like lesions of the adult kidney. Part II: Benign mesenchymal and mixed neoplasms, and tumor-like lesions. *Adv Anat Pathol*. 2000 Jan;7(1):47-66. doi: 10.1097/00125480-200007010-00007
5. Morshid A, Duran ES, Choi WJ, Duran C. A Concise Review of the Multimodality Imaging Features of Renal Cell Carcinoma. *Cureus*. 2021 Feb 8;13(2):e13231. doi: 10.7759/cureus.13231.
6. Metin M, Aydın H, Karaoğlanoğlu M. Renal Cell Carcinoma or Oncocytoma? The Contribution of Diffusion-Weighted Magnetic Resonance Imaging to the Differential Diagnosis of Renal Masses. *Medicina (Kaunas)*. 2022 Feb 1;58(2):221. doi: 10.3390/medicina58020221.
7. Bellin MF, Valente C, Bekdache O, Maxwell F, Balasa C, Savignac A, Meyrignac O. Update on Renal Cell Carcinoma Diagnosis with Novel Imaging Approaches. *Cancers (Basel)*. 2024 May 18;16(10):1926. doi: 10.3390/cancers16101926.
8. Kim JH, Li S, Khandwala Y, Chung KJ, Park HK, Chung BI. Association of Prevalence of Benign Pathologic Findings After Partial Nephrectomy With Preoperative Imaging Patterns in the United States From 2007 to 2014. *JAMA Surg*. 2019 Mar 1;154(3):225-231. doi: 10.1001/jamasurg.2018.4602.
9. Lei Y, Wang H, Li HF, Rao YW, Liu JH, Tian SF, et al. Diagnostic Significance of Diffusion-Weighted MRI in Renal Cancer. *Biomed Res Int*. 2015; 172165. doi: 10.1155/2015/172165.

10. Cova M, Squillaci E, Stacul F, Manenti G, Gava S, Simonetti G, Pozzi-Mucelli R. Diffusion-weighted MRI in the evaluation of renal lesions: preliminary results. *Br J Radiol* 2004. 77(922);851–857. doi: 10.1259/bjr/26525081
11. Sandrasegaran K, Sundaram CP, Ramaswamy R, Akisik FM, Rydberg MP, Lin C, Aisen AM. Usefulness of diffusion-weighted imaging in the evaluation of renal masses. *AJR Am J Roentgenol.* 2010 Feb;194(2):438-45. doi: 10.2214/AJR.09.3024.
12. Doğanay S, Kocakoç E, Çiçekçi M, Ağlamış S, Akpolat N, Orhan I. Ability and utility of diffusion-weighted MRI with different b values in the evaluation of benign and malignant renal lesions. *Clinical Radiology*, 2011. 66(5): 420-425. doi: 10.1016/j.crad.2010.11.013
13. de Silva S, Lockhart KR, Aslan P, Nash P, Hutton A, Malouf D, et al. The diagnostic utility of diffusion weighted MRI imaging and ADC ratio to distinguish benign from malignant renal masses: sorting the kittens from the tigers. *BMC Urol*. 2021 Apr 22;21(1):67. doi: 10.1186/s12894-021-00832-5.
14. Paudyal B, Paudyal P, Tsushima Y, Oriuchi N, Amanuma M, Miyazaki M, et al. The role of the ADC value in the characterisation of renal carcinoma by diffusion-weighted MRI. *Br J Radiol.* 2010 Apr;83(988):336-43. doi: 10.1259/bjr/74949757.
15. Hötker AM, Mazaheri Y, Wibmer A, Zheng J, Moskowitz CS, Tickoo SK, et al. Use of DWI in the Differentiation of Renal Cortical Tumors. *AJR Am J Roentgenol*. 2016 Jan;206(1):100-5. doi: 10.2214/AJR.14.13923.
16. Mahmoud AR, Fouda N, Helmy EM, et al. Role of intravoxel incoherent motion diffusion-weighted MRI in differentiation of renal cell carcinoma subtypes. *Egypt J Radiol Nucl Med.* 55, 184 (2024). doi: 10.1186/s43055-024-01352-6
17. Ding Y, Tan Q, Mao W, Dai C, Hu X, Hou J, et al. Differentiating between malignant and benign renal tumors: do IVIM and diffusion kurtosis imaging perform better than DWI? *Eur Radiol*. 2019 Dec;29(12):6930-6939. doi: 10.1007/s00330-019-06240-6.
18. Ding Y, Zeng M, Rao S, Chen C, Fu C, Zhou J. Comparison of Biexponential and Monoexponential Model of Diffusion-Weighted Imaging for Distinguishing between Common Renal Cell Carcinoma and Fat Poor Angiomyolipoma. *Korean J Radiol.* 2016 Nov-Dec;17(6):853-863. doi: 10.3348/kjr.2016.17.6.853.
19. Panagiotaki E, Chan RW, Dikaios N, Ahmed HU, O'Callaghan J, Freeman A, et al. Microstructural characterization of normal and malignant human prostate tissue with vascular, extracellular, and restricted diffusion for cytometry in tumours magnetic resonance imaging. *Invest Radiol*. 2015 Apr;50(4):218-27. doi: 10.1097/RLI.0000000000000115.
20. Panagiotaki E, Walker-Samuel S, Siow B, Johnson SP, Rajkumar V, Pedley RB, et al. Noninvasive quantification of solid tumor microstructure using VERDICT MRI. *Cancer Res*. 2014 Apr 1;74(7):1902-12. doi: 10.1158/0008-5472.CAN-13-2511.

21. Palombo M, Valindria V, Singh S et al. Joint estimation of relaxation and diffusion tissue parameters for prostate cancer with relaxation-VERDICT MRI. *Sci Rep* 13, 2999 (2023). doi: 10.1038/s41598-023-30182-1.
22. Sen S, Valindria V, Slator PJ, Pye H, Grey A, Freeman A, et al. Differentiating False Positive Lesions from Clinically Significant Cancer and Normal Prostate Tissue Using VERDICT MRI and Other Diffusion Models. *Diagnostics (Basel)*. 2022 Jul 5;12(7):1631. doi: 10.3390/diagnostics12071631.
23. Figini M, Castellano A, Bailo M, Callea M, Cadioli M, Bouyagoub S, et al. Comprehensive Brain Tumour Characterisation with VERDICT-MRI: Evaluation of Cellular and Vascular Measures Validated by Histology. *Cancers (Basel)*. 2023 Apr 27;15(9):2490. doi: 10.3390/cancers15092490.
24. Singh S, Rogers H, Kanber B, Clemente J, Pye H, Johnston EW, et al. Avoiding unnecessary biopsy after multiparametric prostate MRI with VERDICT analysis: The INNOVATE study. *Radiology.* 2022; 305: 623-630. doi: 10.1148/radiol.212536.
25. Johnston EW, Bonet-Carne E, Ferizi U, Yvernault B, Pye H, Patel D, et al. VERDICT MRI for Prostate Cancer: Intracellular Volume Fraction versus Apparent Diffusion Coefficient. *Radiology*. 2019 May;291(2):391-397. doi: 10.1148/radiol.2019181749.
26. Sen S, Singh S, Pye H, et al. ssVERDICT: Self-supervised VERDICT-MRI for enhanced prostate tumor characterization. *Magn Reson Med*. 2024; 92: 2181-2192. doi: 10.1002/mrm.30186
27. Alexander DC. A general framework for experiment design in diffusion mri and its application in measuring direct tissue-microstructure features. *Magn Reson Med*. 2008; 60(2):439–448. doi: 10.1002/mrm.21646
28. Grussu F, Blumberg SB, Battiston M, Kakkar LS, Lin H, Ianus A et al. Feasibility of data-driven, model-free quantitative MRI protocol design: Application to brain and prostate diffusion-relaxation imaging. *Front. Phys.*, 9, November 2021. doi: 10.3389/fphy.2021.752208
29. Blumberg SB, Slator PJ and Alexander DC. Experimental design for multi-channel imaging via task-driven feature selection. In *The Twelfth International Conference on Learning Representations*, 2024. doi: arXiv:2210.06891
30. Panagiotaki E, Ianus A, Johnston E, Chan RW, Stevens N, Atkinson D, et al. Optimised verdict mri protocol for prostate cancer characterisation. In *Proceedings of the International Society for Magnetic Resonance in Medicine (ISMRM)*, 2015. 2872.
31. Veraart J, Fieremans E, Novikov DS. Diffusion MRI noise mapping using random matrix theory. *Magn. Reson. Med*. 2016; 76: 1582-1593. doi: 10.1002/mrm.26059
32. Tournier JD, Smith R, Raffelt D, Tabbara R, Dhollander T, Peitsch M et al. MRtrix3: A fast, flexible and open software framework for medical image processing and

visualisation. *Neuroimage*. 2019; 202: 116137. doi: 10.1016/j.neuroimage.2019.116137

33. Kellner E, Dhital B, Kiselev VG, Reisert M. Gibbs-ringing artifact removal based on local subvoxel-shifts. *Magn. Reson. Med*. 2015; 76: 1574-1581. doi: 10.1002/mrm.26054

34. Smith SM, Jenkinson M, Woolrich MW, Beckmann CF, Behrens TE, Johansen-Berg H, et al. Advances in functional and structural MR image analysis and implementation as FSL. *Neuroimage*. 2004; 23:S208-19. doi: 10.1016/j.neuroimage.2004.07.051

35. Yushkevich PA, Piven J, Hazlett HC et al. 2006. User-guided 3d active contour segmentation of anatomical structures: Significantly improved efficiency and reliability. *Neuroimage.* 31(3):1116-1128. doi: 10.1016/j.neuroimage.2006.01.015

36. Bankhead P, Loughrey MB, Fernández JA et al. QuPath: Open source software for digital pathology image analysis. *Sci Rep*. 2017; 7:16878. doi: 10.1038/s41598-017-17204-5

37. Barbieri S, Gurney-Champion OJ, Klaassen R, Thoeny HC. Deep learning how to fit an intravoxel incoherent motion model to diffusion-weighted MRI. *Magn Reson Med.* 2020 Jan;83(1):312-321. doi: 10.1002/mrm.27910.

38. Virtanen P, Gommers R, Oliphant TE, Haberland M, Reddy T, Cournapeau D, et al. (2020) SciPy 1.0: Fundamental Algorithms for Scientific Computing in Python. *Nature Methods*, 17(3), 261-272. doi: 10.1038/s41592-019-0686-2.

39. Blumberg SB, Lin H, Grussu F, Zhou Y, Figini M, and Alexander DC. Progressive subsampling for oversampled data - application to quantitative MRI. In *MICCAI 2022*, pages 421–431, Cham, 2022. Springer Nature Switzerland. doi: 10.1007/978-3-031-16446-0_40

40. Wang, X., Liu, Q., Kong, W. *et al.* Pathologic analysis of non-neoplastic parenchyma in renal cell carcinoma: a comprehensive observation in radical nephrectomy specimens. *BMC Cancer* 17, 900 (2017). doi: 10.1186/s12885-017-3849-5.